\documentclass[reprint,amssymb,aps,prb,floatfix,usenames,dvipsnames]{revtex4-1}

\usepackage{graphicx}% Include figure files
\usepackage{dcolumn}% Align table columns on decimal point
\usepackage{bm}% bold math
\usepackage[version=4]{mhchem} % chemical formulas
\usepackage{hyperref}% add hypertext capabilities
\usepackage{cleveref} % clever references with \cref command
\usepackage[caption=false]{subfig}
\usepackage{siunitx} % needed for SI units 
\usepackage[colorinlistoftodos]{todonotes} % needed for TODOs

\newcommand{\beq}{\begin{equation}}
\newcommand{\eeq}{\end{equation}}
\newcommand{\mat}[1]{\underline{\underline{#1}}}
\providecommand{\abs}[1]{\lvert#1\rvert}

\newcommand{\new}[1]{\textcolor{black}{#1}}

\Crefname{figure}{Fig.}{.}

\begin{document}

\title{Complex magnetism of B20-MnGe: from spin-spirals, hedgehogs to monopoles}%

\author{Marcel Bornemann}\email{m.bornemann@fz-juelich.de}
\author{Sergii Grytsiuk} 
\author{Paul F. Baumeister} 
\author{Manuel dos Santos Dias} 
\author{Rudolf Zeller} 
\author{Samir Lounis}\email{s.lounis@fz-juelich.de}
\author{Stefan Bl\"ugel}
\affiliation{Peter Gr\"{u}nberg Institut and Institute for Advanced Simulation, Forschungszentrum J\"{u}lich \& JARA, 52425 J\"{u}lich, Germany}

\begin{abstract}
B20 compounds are the playground for various  non-trivial magnetic textures such as  skyrmions, which are topologically protected states. Recent measurements  on B20-MnGe  indicate no clear consensus on its magnetic behavior, which is characterized by the presence of either spin-spirals or  3-dimensional objects interpreted to be a cubic lattice of hedgehogs and anti-hedgehogs. Utilizing a massively parallel linear scaling all-electron density functional algorithm, we find from full first-principles simulations on cells containing thousands of atoms that upon increase of the compound volume, the state with lowest energy switches across different magnetic phases: ferromagnetic, spin-spiral, hedgehog and monopole.
\end{abstract}

\maketitle

\section{Introduction} % Write in your own chapter title
\label{ch:b20}

Most of the current research activity dealing with magnetism in B20 compounds is closely connected to the field of skyrmionics.
Skyrmions are two-dimensional non-trivial magnetization solitons\cite{Bogdanov1989,rosler_spontaneous_2006,everschor-sitte_perspective:_2018}, i.e.\ two-dimensional magnetic structures localized in space, of topological nature\cite{nagaosa_topological_2013}, which have particle-like properties.
Such magnetic objects are heavily prospected with the aim of establishing them as possible information-carrying particles that are small-sized and stable up to room temperature \cite{fert_skyrmions_2013,Tomasello2013,Crum2015,Finocchio2015,Zhang2015b,Dias2016,mruczkiewicz_collective_2016,fert_magnetic_2017,buttner_theory_2018,leroux_skyrmion_2018,LimaFernandes2018,Santos2018}.
This motivated numerous studies on cubic B20-type compounds with broken lattice inversion symmetry\cite{nagaosa_topological_2013,kanazawa_noncentrosymmetric_2017}, where skyrmion phases have been observed for the first time experimentally \cite{muhlbauer_skyrmion_2009}.
\new{\ce{MnGe} is a B20 compound that has been the subject of intense experimental and theoretical investigations, and yet there is still no consensus on how to explain its magnetic properties.
The seminal work of Kanazawa et al.\cite{kanazawa_large_2011} already showed that \ce{MnGe} is rather intriguing.
The magnetic structure was found to have a period between 3 and \SI{6}{\nano\meter}, which is rather short in comparison to the other B20 compounds\cite{kanazawa_noncentrosymmetric_2017}, to be stable up to a temperature of \SI{170}{\kelvin} and a magnetic field of \SI{12}{\tesla}, and to generate a strong topological Hall effect\cite{nagaosa_topological_2013}.
The magnetic structure of \ce{MnGe} has been investigated in reciprocal space via neutron scattering\cite{kanazawa_large_2011,kanazawa_possible_2012,makarova_neutron_2012,grigoriev_chiral_2013,Altynbaev2014,deutsch_two-step_2014,Deutsch2014a,altynbaev_hidden_2016,Martin2017,kanazawa_topological_2017,altynbaev_magnetic_2018,fujishiro_topological_2019}, in real space via Lorentz transmission electron microscopy (LTEM)\cite{shibata_towards_2013,tanigaki_real-space_2015,fujishiro_topological_2019}, and indirectly via transport experiments\cite{kanazawa_large_2011,Shiomi2013,kanazawa_variation_2016,kanazawa_critical_2016}.}

\new{The canonical theory of helimagnetism in B20 compounds\cite{nakanishi_origin_1980,Bak1980,nagaosa_topological_2013} is based on a micromagnetic model featuring two main parameters: the exchange stiffness $A$ and the spin-orbit-driven Dzyaloshinskii-Moriya interaction (or spiralization) $D$.
Several theoretical works have calculated either $D$\cite{koretsune_control_2015,Koretsune2018,mankovsky_composition-dependent_2018,Mankovsky2019} or both $D$ and $A$\cite{gayles_dzyaloshinskii-moriya_2015,kikuchi_dzyaloshinskii-moriya_2016} from density functional theory calculations, but the reported values are quite scattered.
The main finding from these works is that $D$ is too small when combined with $A$ to explain the experimentally observed short period of the magnetic structure, which is given by $\lambda \approx 4\pi A/D$.}

\new{Experimentally, the magnetic structure has been interpreted as consisting of helical spirals\cite{makarova_neutron_2012,Deutsch2014a,altynbaev_hidden_2016,Martin2017,altynbaev_magnetic_2018} as in other B20 compounds, with the short period ascribed to competing long-range magnetic interactions\cite{altynbaev_hidden_2016,altynbaev_magnetic_2018}.
Transport signatures and LTEM imaging strongly back a more complex magnetic structure, which has initially been interpreted as a conventional two-dimensional skyrmion lattice\cite{kanazawa_large_2011,kanazawa_possible_2012,Shiomi2013,shibata_towards_2013}, and afterwards as a three-dimensional skyrmion-antiskyrmion or hedgehog-antihedgehog lattice\cite{tanigaki_real-space_2015,kanazawa_critical_2016,kanazawa_topological_2017,fujishiro_topological_2019}.
Yaouanc and co-workers\cite{yaouanc_magnetic_2017} analyzed muon spin resonance measurements\cite{martin_magnetic_2016} in favor of a helical spiral, and called for further theoretical work to clarify the issue of the magnetic structure of MnGe.
Further evidence favoring the helical structure was provided by recent microwave absorption spectroscopy experiments\cite{turgut_engineering_2018}.}

In this article, we present results on B20-\ce{MnGe} both from a magnetic model approach and large-scale all-electron calculations for a supercell whose extent compares to the periodicity of the experimentally observed helical textures.
After discussing the electronic properties and analyzing the magnetic interactions obtained from the ferromagnetic unit cell, we explore various potential magnetic textures obtained from the self-consistent large-scale ab-initio simulations. \new{The accuracy of the results are scrutinized in terms of convergence of the magnetic interactions as function of the interatomic distances and their sensitivity to various parameters.}
We \new{found} that the experimentally proposed hedgehog structures are marginally higher in energy than the ferromagnetic state around the experimental lattice parameters.
Increasing the volume of the cell can lead to a stabilization of single spirals and of hedgehog lattice states. \new{ While possible disagreement with recent experiments can be blamed on the accuracy of the calculated electronic structure, our investigation calls for further studies on the three-dimensional magnetism in  B20-\ce{MnGe}.}

\section{Methods}
All of the results presented below are \new{based on} Density Functional Theory (DFT) \new{calculations}.
We utilize the \new{DFT} codes juKKR \cite{papanikolaou_conceptual_2002,bauer_development_2014} and KKR\texttt{nano}\cite{thiess_massively_2012} which are both based on the \new{Korringa-Kohn-Rostoker} Green function formalism\cite{papanikolaou_conceptual_2002,Ebert2011a}.
\new{The bilinear magnetic exchange interactions are obtained by infinitesimal rotations of the magnetic moments in the ferromagnetic state\cite{liechtenstein_local_1987,udvardi_first-principles_2003,ebert_anisotropic_2009}, which are implemented in juKKR following Ref.~\onlinecite{ebert_anisotropic_2009}.}
KKR\texttt{nano} was especially designed to perform large-scale electronic structure calculations and allows us to perform self-consistent all-electron calculations for supercells that contain a few thousand atoms \cite{thiess_massively_2012}.
\new{The finite-temperature energy integration method of Wildberger et al.\ was used\cite{wildberger_fermi-dirac_1995} with $T=\SI{800}{\kelvin}$.}

In \cref{sec:basic_prop_mnge}, we review the basic electronic and magnetic properties of the primitive cell of B20-\ce{MnGe}.
To \new{identify} the appropriate exchange and correlation functional to utilize for the large supercells, we \new{determine} the theoretical equilibrium lattice parameter within the scalar-relativistic approximation as implemented in KKR\texttt{nano}.
As exchange-correlation functionals, we choose the local density approximation (LDA) according to the spin-dependent scheme of Vosko, Wilk and Nusair \cite{vosko_accurate_1980}, and the generalized gradient approximation as given in PBEsol \cite{perdew_restoring_2008}.
A grid of $14\times14\times14$ k-points is used.

In \cref{{sec:micro_par}}, we \new{perform first-principles calculations including spin-orbit coupling\cite{bauer_development_2014} with the} juKKR code to extract the \new{bilinear magnetic pair interactions between the \ce{Mn} atoms in B20-MnGe, namely} \new{the} isotropic exchange interactions ($J_{ij}$) and the
Dzyaloshinskii-Moriya interactions ($\vec{D}_{ij}$).
The k-point mesh was increased to $60\times60\times60$ points.
\new{Two kinds of calculations are performed: (i) at the theoretical PBEsol lattice parameter but displacing the internal positions of the Mn atoms, and (ii) keeping the internal positions of the Mn atoms fixed but varying the lattice parameter.}
%\old{The calculations are performed for $a=\SI{4.76}{\angstrom}$, i.e.\ the theoretical lattice parameter obtained with PBEsol and larger lattice parameters.
%In order to also investigate the impact of a temperature-driven small shift of atomic positions, as reported from a neutron diffraction study\cite{makarova_neutron_2012}, the interaction parameters are determined for the experimentally obtained positions of the \ce{Mn} atoms with the atomic positioning parameter $u_{\ce{Mn}}=0.135$ and a slightly smaller (larger) value of $u_{\ce{Mn}}=0.125$ ($u_{\ce{Mn}}=0.145$).
%When the \ce{Mn} atoms are moved, the distance between \ce{Mn} and \ce{Ge} atoms in the crystal structure changes.}
Additionally, we use an in-house Monte-Carlo code that only considers the isotropic exchange interactions to determine the Curie temperature of the system \new{using the Metropolis algorithm}\cite{PhysRevB.88.134403,Matsumoto:1998:MTE:272991.272995}.
\new{We also performed atomistic spin dynamics simulations with the Spirit code\cite{noauthor_spirit_nodate}, which are based on the numerical solution of the Landau-Lifshitz-Gilbert equation for a spin lattice model that takes into account all the computed bilinear magnetic interactions, and that can find the magnetic ground state and also metastable states.}

For the large-scale calculations with KKR\texttt{nano} in \cref{sec:ls_KKRnano_calculations} we use supercells built of $6\times6\times6$ unit cells so that 1728 atoms are treated \new{including spin-orbit coupling}.
PBEsol is used as exchange-correlation functional and only a single $k$-point, i.e. the $\Gamma$-point, is included.
The Green function is truncated beyond a distance of $2a$.
The magnetic states are imposed on the system by forcing the atomic exchange-correlation B-fields to point into specific directions.

\section{Basic Properties of ferromagnetic B20-M\lowercase{n}G\lowercase{e}}
\label{sec:basic_prop_mnge}
\begin{figure}[tb]
  \centering
   \includegraphics[width=0.8\columnwidth]{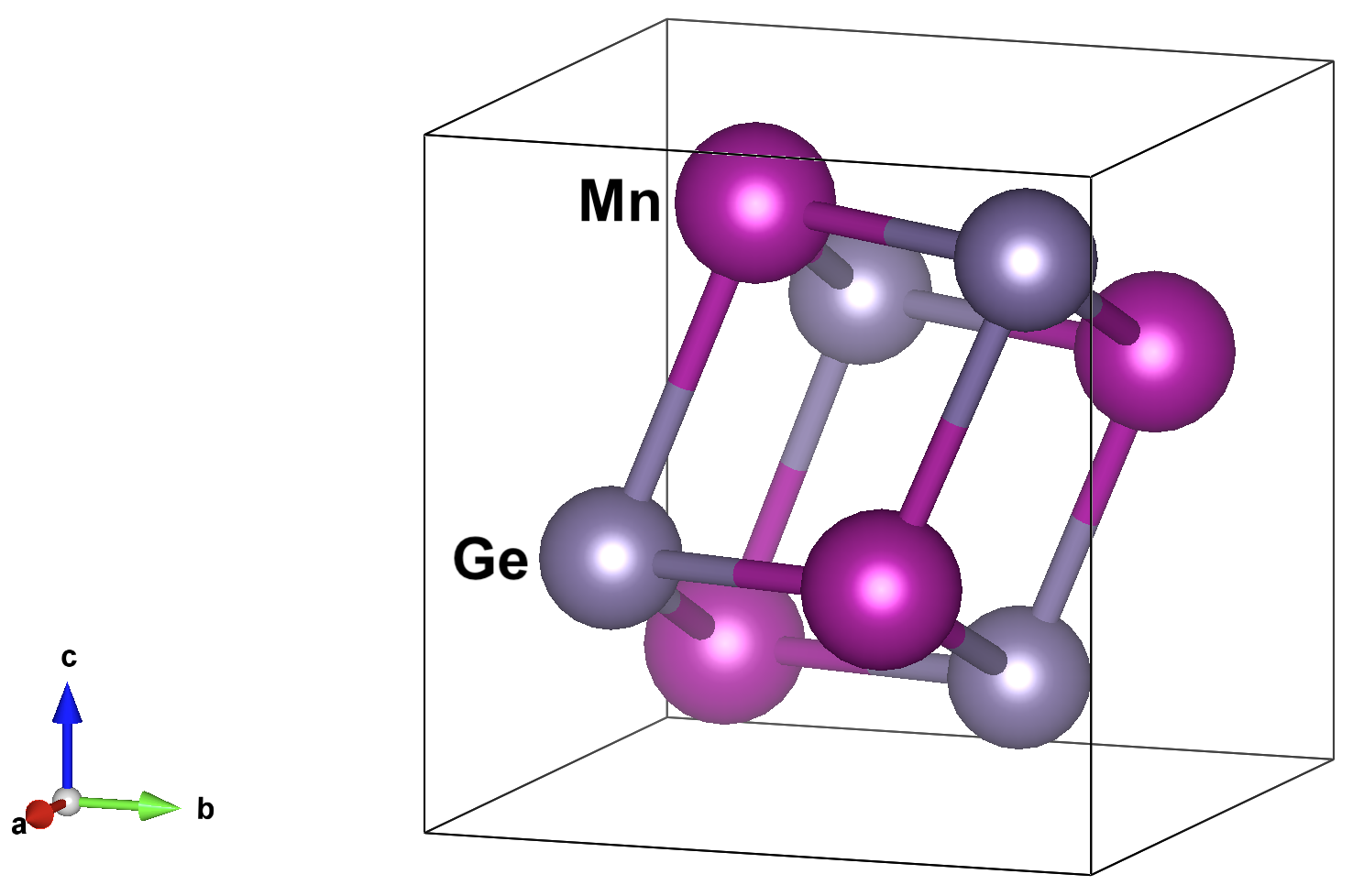}
	\caption{Crystal structure of B20-\ce{MnGe}.
	The four Mn atoms are equivalent, as are the four Ge atoms.
	}
\label{fig:MnGe_struct}
\end{figure}

B20-\ce{MnGe} orders in a cubic structure that is described by the $P2_{1}3$ space group.
This space group is noncentrosymmetric, which means that there is no lattice inversion symmetry.
The eight atoms in the primitive cell are located at the Wyckoff positions $(u,u,u)$, $(1/2-u,1-u,1/2+ u)$, $(1-u,1/2+u,1/2-u)$ and $(1/2+u, 1/2-u, 1-u)$, where $u$ is a constant value that is determined for each atom type.
We choose these parameters as $u_{\ce{Mn}}=0.135$  and $u_{\ce{Ge}}=0.8435$ which is in good agreement with experimental findings\cite{makarova_neutron_2012,ditusa_magnetic_2014,Dyadkin2014}.
\new{The crystal structure is represented in Fig.~\ref{fig:MnGe_struct}.}

\begin{figure}[tb]
  \centering
   \includegraphics[width=1.0\columnwidth]{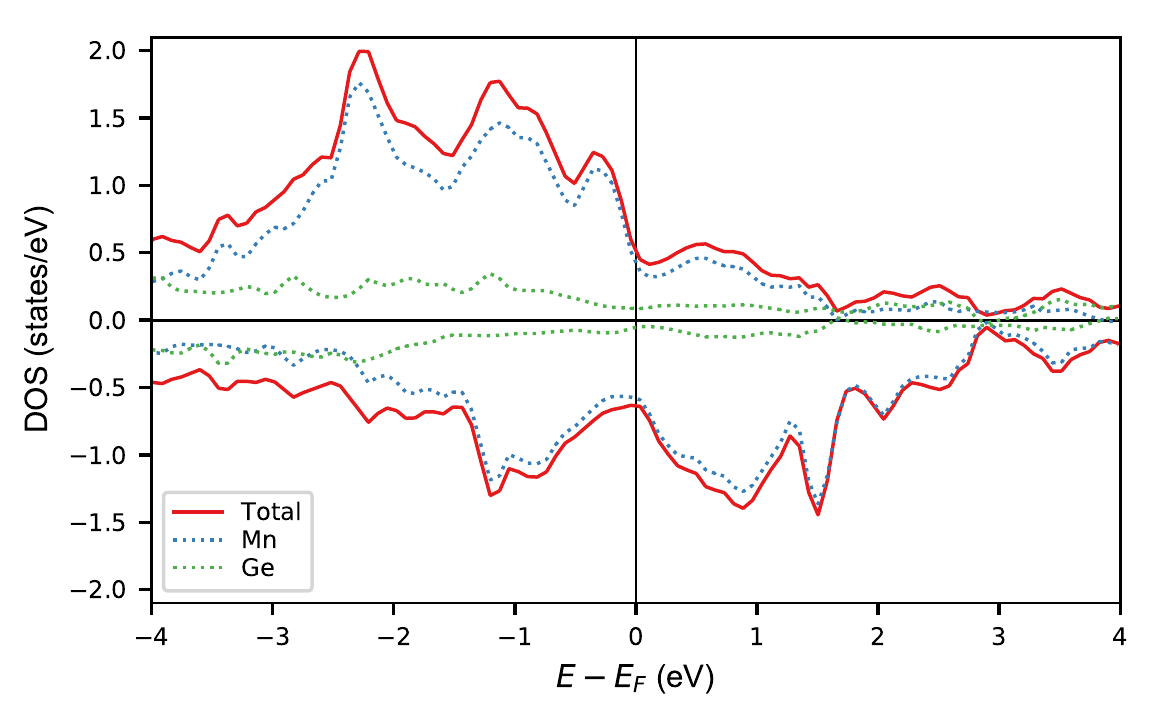}
	\caption{Energy-resolved density of states for B20-\ce{MnGe}.
	The contributions of the \ce{Mn} bands and the \ce{Ge} bands
	are plotted separately with dotted lines while
	the combined density of both the \ce{Mn} and the \ce{Ge} states is indicated by
	a solid line.
	The majority (minority) spin channel is 
	denoted by positive (negative) y-values. The spin splitting, 
	which gives rise to the magnetic moment of
	the \ce{Mn} atoms, is clearly recognizable.
	}
\label{fig:MnGe_dos}
\end{figure}

\new{We start by computing the basic electronic properties of the ferromagnetic 8-atom unit cell with the lattice parameter set to the experimental lattice parameter, $a=\SI{4.79}{\angstrom}$\cite{makarova_neutron_2012,ditusa_magnetic_2014,Dyadkin2014}.}
In \Cref{fig:MnGe_dos} the density of states obtained with KKR\texttt{nano} is shown.
One recognizes the large spin-splitting characterizing Mn atoms, which gives rise to a magnetic moment of roughly $2\,\mu_\mathrm{B}$ per Mn atom.
The \ce{Ge} states do not contribute significantly to the density of states at and around the Fermi level.

\begin{figure}[tb]
  \centering
   \includegraphics[width=1.0\columnwidth]{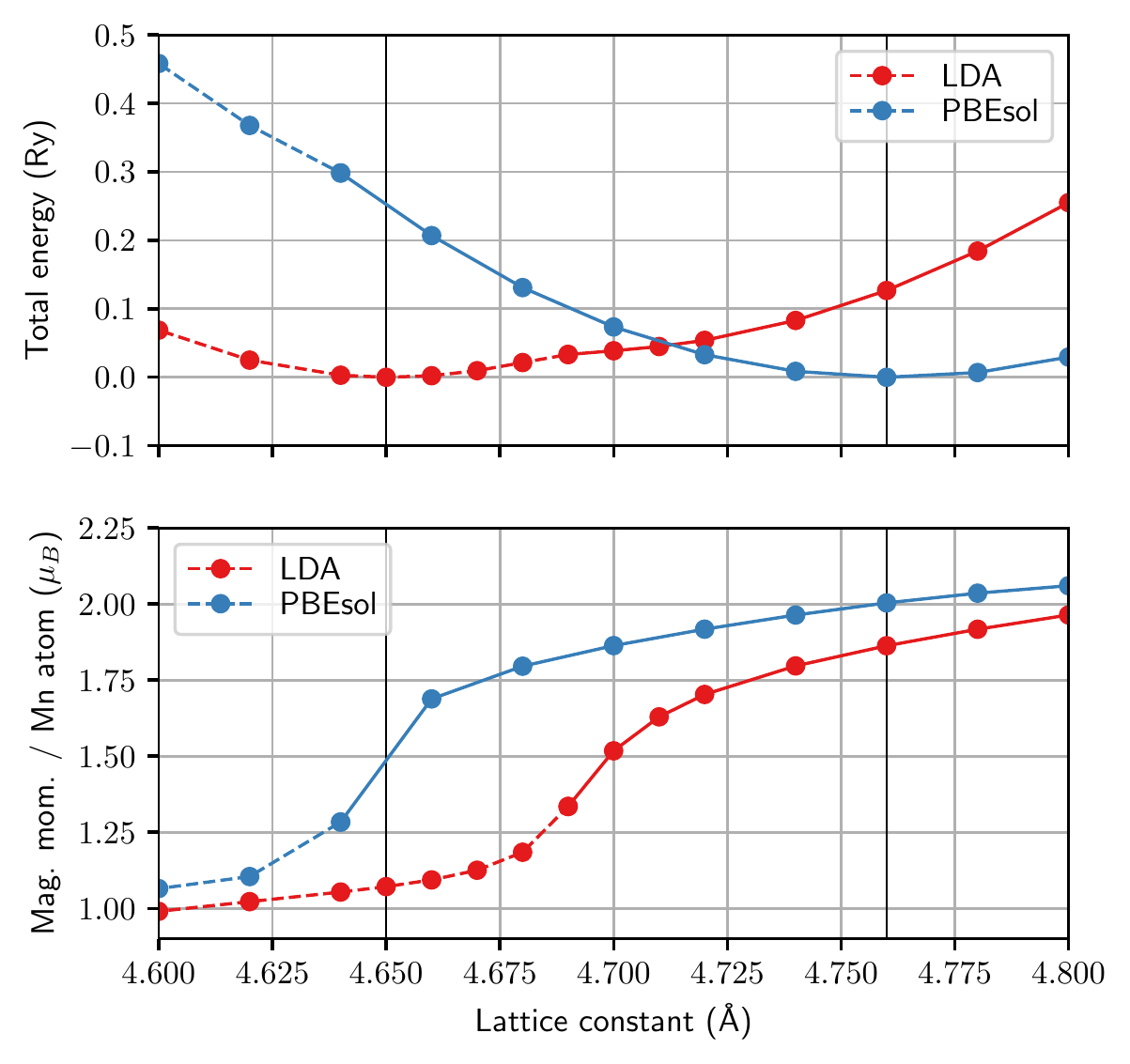}
   \caption{Comparison of LDA and PBEsol exchange-correlation functionals for B20-\ce{MnGe}.
        Top: Total energy vs.\ lattice parameter with LDA and PBEsol functionals.
        \new{The equilibrium lattice constants for each functional are marked by two vertical black lines.}
        %For LDA the equilibrium lattice constant is found at $a = \SI{4.65}{\angstrom}$, while PBEsol predicts $a = \SI{4.76}{\angstrom}$ \new{(these values are marked by the two vertical black lines)}.
	    The respective minimum energy is taken as energy zero for the corresponding total energy curve.
   		Bottom: Magnetic moment per Mn atom vs.\ lattice parameter, \new{showing a high-spin (large magnetic moment) to low-spin (small magnetic moment) transition for lattice compression.}
   		%If pressure is applied, both LDA and PBEsol predict a transition from a high-spin state (solid line on the right-hand side of the figure), in which the magnetic moment per Mn atom is around $2\,\mu_\mathrm{B}$, to a low-spin state (dashed line on the left-hand side of the figure), where the moment is about $1\,\mu_\mathrm{B}$.
		%While LDA places the crossover region near $a = \SI{4.7}{\angstrom}$, with PBEsol it is at $a \approx \SI{4.65}{\angstrom}$.
		}
\label{fig:MnGe_hsls}
\end{figure}

\new{Next we explore how the magnetic moment depends on the assumed lattice constant.}
We recover the pressure-induced magnetic transition from a high-spin state (HS) to a low-spin (LS) state (see \Cref{fig:MnGe_hsls}), \new{i.e.\ a significant and rather abrupt reduction of the magnetic moment of \ce{Mn}, that was} predicted by R\"o{\ss}ler \cite{rosler_ab_2012} and confirmed experimentally by Deutsch et al.\cite{deutsch_two-step_2014}.
The latter reported additionally that the helical ordering in the material collapses above an applied pressure of \SI{10}{\giga\pascal}. 

\new{The computed total energies as a function of the lattice parameter show that the LDA equilibrium lattice constant, $a = \SI{4.65}{\angstrom}$, is substantially lower than the PBEsol one, $a = \SI{4.76}{\angstrom}$, see \Cref{fig:MnGe_hsls}, and that the latter is closer to the experimental value, $a = \SI{4.79}{\angstrom}$ \cite{makarova_neutron_2012,ditusa_magnetic_2014,Dyadkin2014}.}
The calculation also sheds light on the behavior of the magnetic moment under variation of the
lattice constant.
As can be seen in the lower part of \Cref{fig:MnGe_hsls}, the magnetic moment of each \ce{Mn} atom becomes larger
with increasing lattice constant.
The main difference between LDA and PBEsol is the location of the crossover region between the LS and HS state in which the moments
increase abruptly and the system goes into the HS state.
For PBEsol it is found around $a = \SI{4.65}{\angstrom}$ while it is slightly below $a = \SI{4.7}{\angstrom}$ for LDA.
% Interestingly, and as observed in previous works\cite{rosler_ab_2012}, the LDA equilibrium lattice is associated with the LS in contrast to PBEsol (HS). 
Furthermore, it is remarkable that the magnetic moments per \ce{Mn} atom differ a lot for the equilibrium
lattice constant of LDA and PBEsol. Here, LDA predicts a magnetic moment \new{per Mn atom} that is a bit larger than
$1\,\mu_\mathrm{B}$, where instead PBEsol \new{yields} a value of almost $2\,\mu_\mathrm{B}$.
Experimentally, the magnetic moment was reported in the range $1.6 - 2.3\,\mu_\mathrm{B}$\cite{kanazawa_large_2011,makarova_neutron_2012,deutsch_two-step_2014,Deutsch2014a}.
A closer look reveals that there are actually two parabola-like energy curves for each functional.
One describes the total energy of the system in the HS state (solid line)
while the other does the same for the LS state (dashed line).
At the transition point the two curves intersect and the two states are degenerate.

To summarize, a HS/LS transition is predicted with both LDA and PBEsol, where PBEsol correctly finds the ground state to be the HS state while LDA does not\new{, for the respective theoretical lattice parameters}.
\new{As the PBEsol equilibrium lattice constant is also much closer to the experimental value}, we restrict ourselves to PBEsol as exchange-correlation functional in the following.
%when investigating the large cells containing more than a thousand magnetic atoms.

\section{Magnetic exchange interactions}
\label{sec:micro_par}

\new{The magnetic pair interactions are associated with the following extended Heisenberg model
\beq
\label{eq:heisenberg_hamiltonian}
E_{\text{atom}} = -\sum_{ij} J_{ij}\,\vec{m}_i\cdot\vec{m}_j +
\sum_{ij} \vec{D}_{ij} \cdot \left(\vec{m}_i \times \vec{m}_j \right),
\eeq
where $\vec{m}_i$ is the unit vector for the orientation of the Mn magnetic moment at site $i$,  $J_{ij}$ is the isotropic exchange interaction and $\vec{D}_{ij}$ is the vector characterizing the DMI, and both connect the moments on sites $i$ and $j$.
Since each \ce{Ge} atom carries a small magnetic moment of $0.1\,\mu_\mathrm{B}$, they are not considered in our magnetic models.}

\new{The canonical theory of helical magnetic structures in the B20 materials is based on the competition between the isotropic exchange interactions and the DMI\cite{nakanishi_origin_1980,Bak1980,nagaosa_topological_2013}.
As the helical period in these materials tends to be much larger than the lattice parameter, this theory is conveniently expressed by a micromagnetic model where the energy reads
\begin{align}\label{eq:micro_hamiltonian}
  E_{\text{micro}} =& \int \mathrm{d}V\, [ A \left((\nabla m_x)^2 +  (\nabla m_y)^2 + (\nabla m_z)^2 \right) \nonumber \\
&+ D\,\vec{m} \cdot \left(\nabla \times \vec{m} \right) ],
\end{align}
where $\vec{m}(\vec{r}\,)$ is the normalized magnetization field, $A$ is the exchange stiffness, and $D$ is the DM spiralization as already mentioned in the introduction.
The period of the helical structure $\lambda$ is then proportional to $A/D$ and given by the provision
\begin{equation}
    \label{eq:hel_period}
     \lambda = \frac{2\pi}{q} = -4\pi\frac{A}{D},
\end{equation}
where $q$ is the wave number.
Note that this expression is only valid for the helical 1Q phase.
In order to apply this theory to MnGe, the exchange stiffness $A$ and the DM spiralization $D$ are evaluated from their respective interatomic counterparts, $J_{ij}$ and $\vec{D}_{ij}$, respectively, by summation over all pairs of magnetic sites up to a chosen cutoff distance.
The relation between the lattice and micromagnetic models is discussed in \cref{sec:con_atom_micro}.}

\new{In this section, we first compute the magnetic pair interactions from the ferromagnetic state and then construct the micromagnetic parameters $A$ and $D$.}

\subsection{Magnetic pair interactions}
\begin{figure*}[htb]
	\centering
	\includegraphics[width=0.80\textwidth]{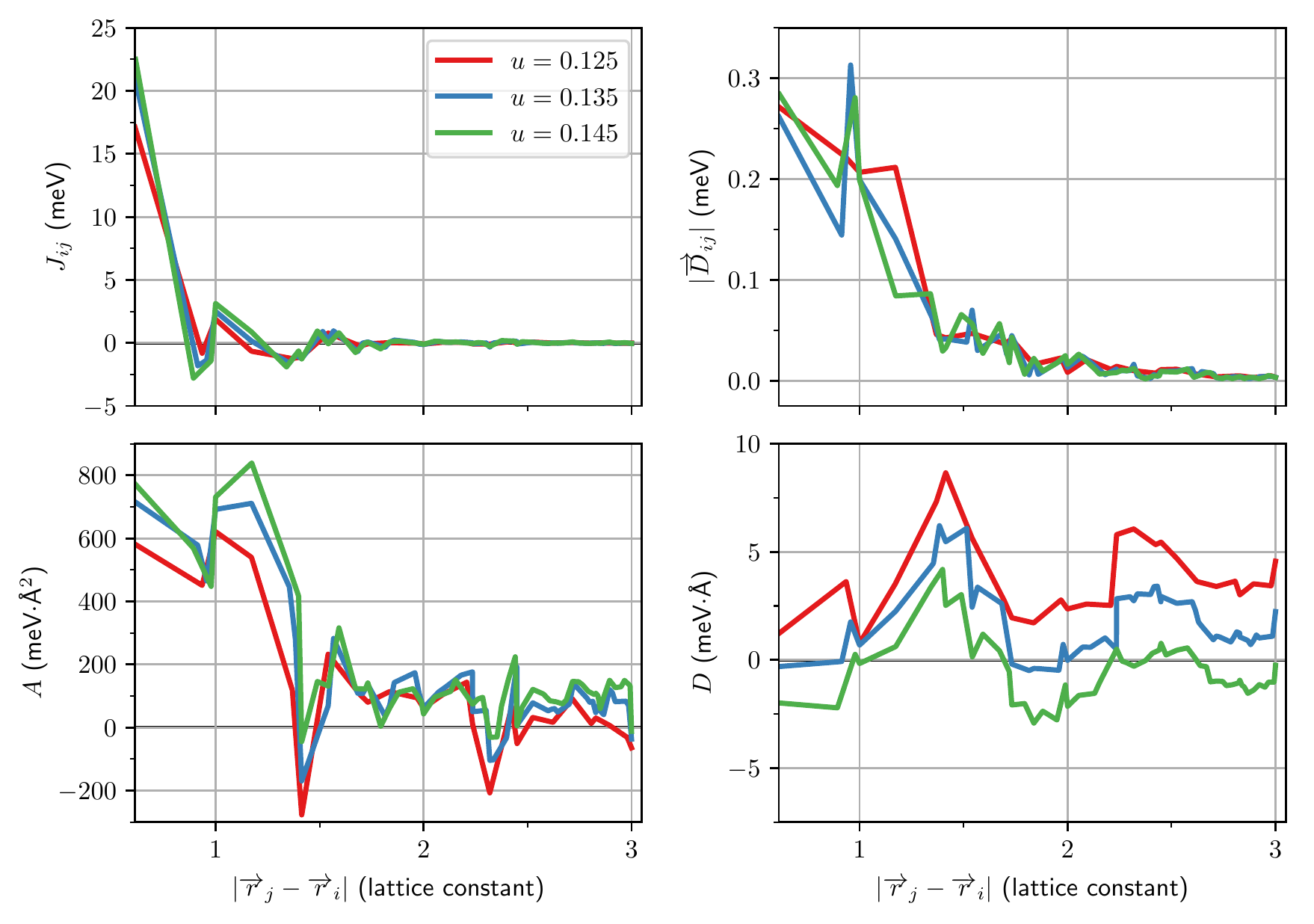}
    \caption{\new{Dependence of the magnetic interactions on the interatomic distance between Mn atoms when the internal Mn positions are varied.
        The \textit{ab initio} calculations were performed with the juKKR code using the PBEsol functional at its theoretical lattice parameter, $a = \SI{4.76}{\angstrom}$.
		The internal positions of the Mn atoms were varied by setting $u_{\ce{Mn}} = 0.125,\,0.135$ and $0.145$.
		Top panels: isotropic magnetic exchange couplings $J_{ij}$ and absolute values of the DMI vectors $\abs{\vec{D}_{ij}}$ as a function of the interatomic distance between Mn atoms.
		Bottom panels: micromagnetic spin stiffness $A$ and micromagnetic DM spiralization $D$ computed from the magnetic pair interactions by summation up to the given interatomic distance.
		%For the next nearest neighbor the isotropic exchange interaction is considerably larger than the DMI. For more remote atomic pairs the difference in magnitude between the two decreases. 
		%The sign of $J_{ij}$ alternates with distance giving rise to magnetic frustration as a positive (negative) value corresponds to (anti-)ferromagnetic coupling between two atoms.
		$A$ and $D$ do not converge to a constant value with the inclusion of further shells in the summations.}}
 \label{fig:mnge_varied_u} 
\end{figure*}
\new{We investigate the sensitivity of the computed magnetic interactions to the structural parameters defining the unit cell in two ways.
First, we keep the lattice parameter fixed at the theoretical PBEsol value, but vary the internal positions of the Mn atoms by setting $u_{\ce{Mn}} = 0.125,\,0.135$ and $0.145$ (the positions of the Ge atoms are fixed).
The computed $J_{ij}$ and $\abs{\vec{D}_{ij}}$ are shown in the top two plots of \Cref{fig:mnge_varied_u}.
We call each distinct interatomic separation between Mn atoms a shell.
The first $J_{ij}$ shell (nearest-neighbor interaction) is always large and positive, indicating strong ferromagnetic coupling.
The next $J_{ij}$ shells alternate in sign, which indicates magnetic frustration, and are more sensitive to the $u$-parameter.
The values of $\abs{\vec{D}_{ij}}$ are two orders of magnitude smaller than the values of $J_{ij}$, and shells other than the first also show sensitivity to the choice of $u$-parameter, as was the case for the isotropic interactions.
We rationalize this behavior as follows:
When the internal positions of the \ce{Mn} atoms are varied, the distance between pairs of \ce{Mn} and \ce{Ge} atoms in the crystal structure changes.
This modifies the hybridization between both atom types for certain pairs, which might be the reason for the found changes in the interactions.}
%The second shell is small and negative, and weakens with decreasing $u$-parameter.
%A detailed account of the magnetic exchange interactions is given in \Cref{fig:mnge_varied_u}, where both the atomistic model parameters $J_{ij}$ and $\abs{\vec{D}_{ij}}$ are depicted for $u_{\ce{Mn}}=0.125$, $u_{\ce{Mn}}=0.135$ and $u_{\ce{Mn}}=0.145$.
%For all three $u$-parameters the first nearest-neighbor interaction is strongly positive which means that ferromagnetic coupling is preferred here.
%However, the alternating sign of the second nearest-neighbor isotropic exchange interactions indicates magnetic frustration.
%Interestingly, a different choice of the structural parameter $u$ has a strong influence on their behavior between the second and fifth shell.

\begin{figure*}[htb]
	\centering
	\includegraphics[width=0.80\textwidth]{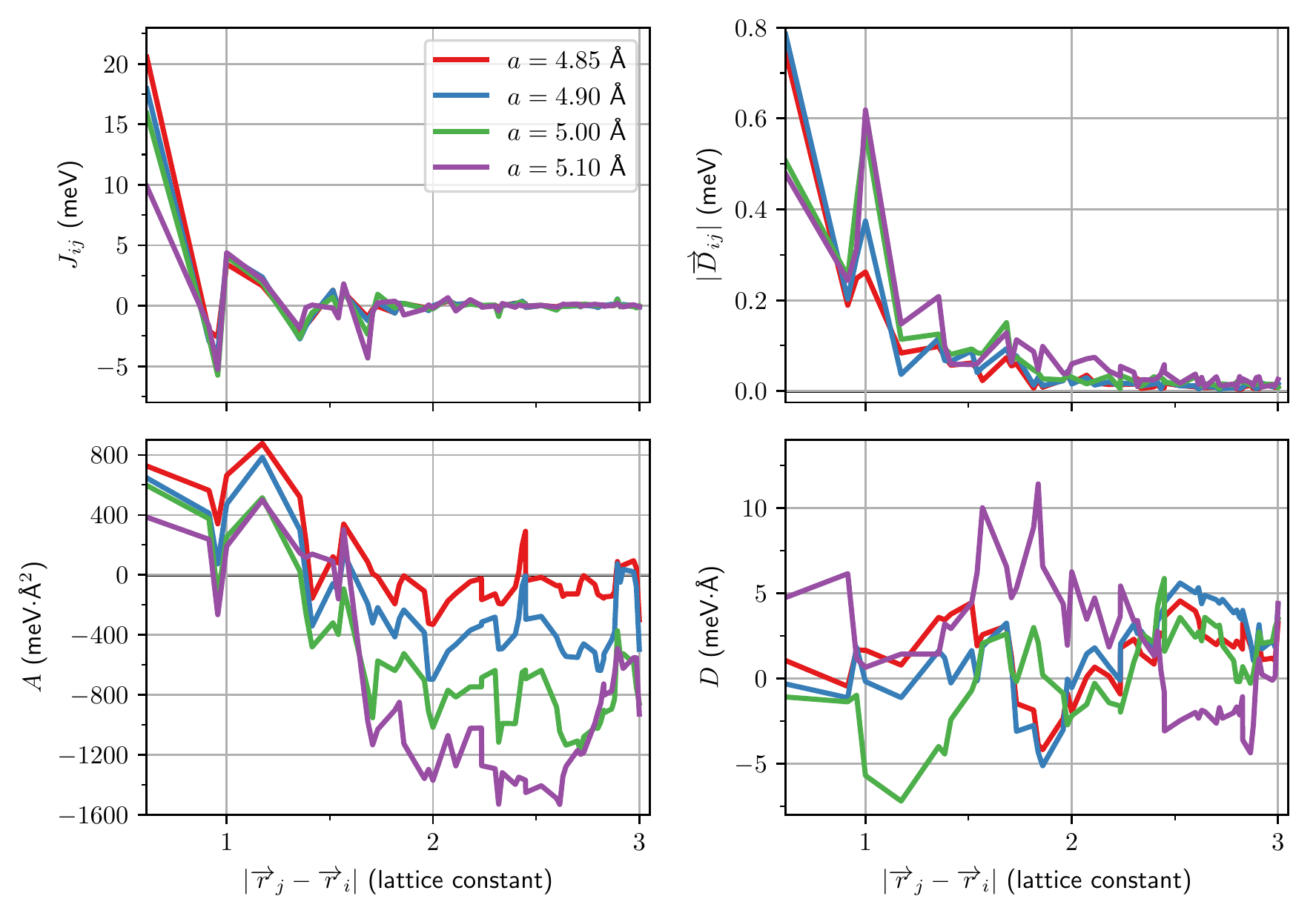}
	\caption{\new{Dependence of the magnetic interactions on the interatomic distance between Mn atoms when the lattice parameter is varied.
	    The \textit{ab initio} calculations were performed with the juKKR code using the PBEsol functional at fixed $u_{\ce{Mn}}=0.135$.
	    The lattice parameters were set to $a = 4.85, 4.90, 5.00$ and \SI{5.10}{\angstrom}.
	    Top panels: isotropic magnetic exchange couplings $J_{ij}$ and absolute values of the DMI vectors $\abs{\vec{D}_{ij}}$ as a function of the interatomic distance between Mn atoms.
		Bottom panels: micromagnetic spin stiffness $A$ and micromagnetic DM spiralization $D$ computed from the magnetic pair interactions by summation up to the given interatomic distance.
	    %Isotropic magnetic exchange couplings ($J_{ij}$) and absolute values of the DMI vectors ($\left|{\vec{D_{ij}}}\right|$) as function of distance between atoms $i$ and $j$ derived from an ab initio calculation with KKR using the PBEsol functional.
		%The colors indicate the chosen lattice parameter.
		%For the next nearest neighbor the isotropic exchange interaction is considerably larger than the DMI. For more remote atomic pairs the difference in magnitude between the two decreases. 
		%The sign of $J_{ij}$ alternates with distance giving rise to magnetic frustration as a positive (negative) value corresponds to (anti-)ferromagnetic coupling between two atoms.
		$A$ and $D$ do not converge to a constant value with the inclusion of further shells in the summations.}}
 \label{fig:mnge_varied_alat} 
\end{figure*}
\new{Second, we keep the internal positions fixed ($u_{\ce{Mn}}=0.135$) and vary the lattice parameter.
We consider only lattice expansion, as we expect that it might weaken the dominant nearest-neighbor ferromagnetic coupling, and so open the possibility of a different magnetic ground state.
The values of $J_{ij}$ and $\abs{\vec{D}_{ij}}$ are shown in \Cref{fig:mnge_varied_alat} for $a = 4.85, 4.90, 5.00$ and \SI{5.10}{\angstrom}, which are all larger than the theoretical PBEsol lattice constant ($a = \SI{4.76}{\angstrom}$).
As anticipated, the $J_{ij}$ for the first shell are strongly reduced with increasing lattice parameter, while the ones for further shells are much less affected.
The behavior of $\abs{\vec{D}_{ij}}$ differs between the two smaller and the two larger lattice parameters.
For $a=4.85$ and \SI{4.90}{\angstrom}, the $\abs{\vec{D}_{ij}}$ for the first shell is the largest, while for $a=5.00$ and \SI{5.10}{\angstrom} it is strongly reduced while the $\abs{\vec{D}_{ij}}$ for the fourth shell is strongly enhanced and now dominates over the one of the first shell.}
%Another structural parameter whose effect on the magnetic model parameters can be studied is the lattice constant.
%Our results are shown in \Cref{fig:mnge_varied_alat}, where the atomistic and micromagnetic parameters for $a=4.85, 4.90, 5.00$ and $5.10$ \AA \, are visualized.
%All of these lie above the equilibrium lattice constant.
%By increasing $a$, and therefore the inter-atomic distances, the first nearest-neighbor isotropic exchange can be reduced.
%The $J_{ij}$-couplings between the neighbors that lay further apart are less affected.
%Interestingly, the behavior changes for the larger lattice constants.
%Here, the most significant $\vec{D}_{ij}$ values are found for $\left| \vec{r}_{j} - \vec{r}_{i}\right| = 1.0$ $a$.

\subsection{Micromagnetic parameters}
%Since the magnetic behavior of MnGe is often described in terms of a micromagnetic model (see e.g. Ref.\cite{nagaosa_topological_2013}) where the energy reads
%\begin{align}
%E_{\text{micro}} =& \int dV\, [ A \left((\nabla m_x)^2 +  (\nabla m_y)^2 + (\nabla m_z)^2 \right) \nonumber \\
%&+ D \vec{m} \cdot \left(\nabla \times \vec{m} \right) ],
%\end{align}
%it is useful to evaluate the exchange stiffness, $A$, and the DM spiralization $D$ from their respective interatomic counterparts,  $J_{ij}$ and $D_{ij}$
%by summation over individual shells
%(see the thorough derivation in \cref{sec:con_atom_micro}).
%The helical period of a magnetic structure is then proportional to $A/D$ and given by the provision
%\begin{equation}
%    \label{eq:hel_period}
%     \lambda = \frac{2\pi}{q}  = -4\pi\frac{A}{D},
%\end{equation}
%where $q$ is the wave number of the spin spiral.
%Note, that this expression is only valid for the helical 1Q phase.
\new{Now we turn to the micromagnetic spin stiffness $A$ and DM spiralization $D$, which are defined in terms of their interatomic counterparts as explained in \cref{sec:con_atom_micro}.
The results are shown in the bottom panels of \Cref{fig:mnge_varied_u} and \Cref{fig:mnge_varied_alat}.
The first observation is that the micromagnetic parameters do not converge with increased interatomic cutoff distance in the summations.
This prevents us from evaluating the helical pitch $\lambda$ by \cref{eq:hel_period}, according to the canonical theory.
Note that estimating the pitch in this manner is only valid if the magnetic texture is truly created by the competition of the spin stiffness and DM spiralization as given by the micromagnetic model.
If frustration of the magnetic interactions plays a crucial role\cite{leonov_multiply_2015}, this theory is no longer applicable.
On the other hand, this poor convergence indicates that the underlying magnetic pair interactions are extremely long-ranged.
This agrees with an alternative scenario for the origin of the magnetic ground state of MnGe, which is based on RKKY interactions\cite{altynbaev_hidden_2016,altynbaev_magnetic_2018}.}

\new{Despite the lack of convergence of the summations, many conclusions can still be taken from studying the micromagnetic parameters.
First we discuss the role played by varying the internal positions of the Mn atoms.
As a function of the interatomic cutoff distance in the summations, the spin stiffness starts from fairly large values, dips at $\approx 0.9a$ due to antiferromagnetic $J_{ij}$ (c.f.\ top left panel of \Cref{fig:mnge_varied_u}), before plunging to fairly small ones when the cutoff exceeds $\approx 1.5a$ (see bottom left panel of \Cref{fig:mnge_varied_u}).
This is both an indication that interactions between far-away atoms are important, sustaining our claim that they are long-ranged, but also that the interactions have a competing nature, so that when they are summed up the resulting value is low.
While the values of $\abs{\vec{D}_{ij}}$ for the first shell were quite insensitive to changes in $u_{\ce{Mn}}$, the same is not true for the micromagnetic $D$ computed from those pairwise interactions (see first three data points on bottom right panel of \Cref{fig:mnge_varied_u}).
%While the absolute values of $\vec{D}_{ij}$ are very similar for the first shell as function of the $u$-parameter, the values of $D$ differ if only the first shell is taken into account.
This results from the progressive rotation of the DMI vectors $\vec{D}_{ij}$ with respect to the bond vector connecting the two Mn atoms, which is central to the formula in \cref{eq:spir_tensor} that defines $D$, and explains why the value of $D$ changes from positive to near-zero to negative as $u_{\ce{Mn}}$ increases.
This decrease of $D$ when $u_{\ce{Mn}}$ increases is actually a general trend, as seen by the relative ordering of the three computed curves.}

\new{Next we discuss the impact of expanding the lattice on the micromagnetic parameters.
As a function of the cutoff distance in the summations, the spin stiffness initially follows the same trend as the first shell of the $J_{ij}$, weakening as the lattice parameter is increased (see the bottom panel of \Cref{fig:mnge_varied_alat}).
There is a strong dip at $\approx 0.9a$ due to antiferromagnetic $J_{ij}$ (c.f.\ top left panel of \Cref{fig:mnge_varied_alat}), but then $A$ increases again to a large positive value.
Strikingly, an abrupt change once again takes place when the cutoff exceeds $\approx 1.5a$: $A$ becomes negative, and its trend is to become more negative the more the lattice is expanded.
The trigger is another shell of strong antiferromagnetic $J_{ij}$, but the remaining long-range part of the pair interactions does not seem to have a definite sign, so the overall negative tendency remains for larger cutoff distances.
A negative spin stiffness immediately invalidates the assumptions behind the micromagnetic model of \cref{eq:micro_hamiltonian}, namely that the interactions are dominantly ferromagnetic, and so $A > 0$.
This could be a hint to the existence of a helical magnetic texture driven by competing isotropic exchange interactions, instead of DMI.
The spiralization tensor also shows a strong evolution with increasing lattice parameter.
Starting again from the first shell of $\vec{D}_{ij}$, we already found that when the lattice parameter increases their magnitude decreases, but the micromagnetic $D$ has a completely different behavior, first staying at fairly low values before increasing drastically for the largest lattice parameter considered, due to a realignment of the $\vec{D}_{ij}$ with the bond vectors connecting the atoms.
Increasing the cutoff distance in the summations does not produce any discernible trend in the behavior of $D$, though.}

\new{We close the discussion of the micromagnetic parameters by placing our results in the context of previous works on MnGe.
Informed by the canonical micromagnetic theory, most attention has focused on the $D$ parameter.
An approach by Koretsune et al.\ based on taking the limit of the static non-uniform spin susceptibility constructed from a Wannier representation resulted in the value $D = \SI{107}{\milli\electronvolt\angstrom}$\cite{koretsune_control_2015}, with a revised value of $D \approx \SI{27}{\milli\electronvolt\angstrom}$ computed by the same method being reported in a recent review\cite{Koretsune2018}.
Different approaches, based on the DFT energies of spin spirals or a different derivation of the spiralization tensor, resulted in the values $D = \SI{1.2}{\milli\electronvolt\angstrom}$\cite{gayles_dzyaloshinskii-moriya_2015}, $D \approx \SI{-1.8}{\milli\electronvolt\angstrom}$\cite{kikuchi_dzyaloshinskii-moriya_2016}, and $D \approx \SI{1}{\milli\electronvolt\angstrom}$\cite{mankovsky_composition-dependent_2018,Mankovsky2019}.
We see that the literature consensus is for a small value of $D$, of the order of magnitude of the values we report on the bottom right panel of \Cref{fig:mnge_varied_u}, with the results of the susceptibility method unexplained outliers.
There seems to be less interest in the literature regarding the spin stiffness, with only two reported values: $A = \SI{280}{\milli\electronvolt\angstrom\square}$\cite{gayles_dzyaloshinskii-moriya_2015} and $A \approx \SI{800}{\milli\electronvolt\angstrom\square}$\cite{kikuchi_dzyaloshinskii-moriya_2016}.
The first value is an upper bound for our computed stiffness at large cutoff distances, while the second value is similar to our computed stiffness from the first couple of shells of $J_{ij}$ interactions, see bottom left panel of \Cref{fig:mnge_varied_u}.}

\subsection{Atomistic spin dynamics}
\new{The preceding analysis of the trends and properties of the micromagnetic parameters raised some doubts on whether the canonical theory based on the micromagnetic model is the right approach to describe the magnetism of MnGe.
However, we can revert to the lattice spin model of \cref{eq:heisenberg_hamiltonian} and explore its output.} 
%Samir see if it provides a better picture.}

\begin{figure}[tb]
	\centering
	\includegraphics[width=1.0\columnwidth]{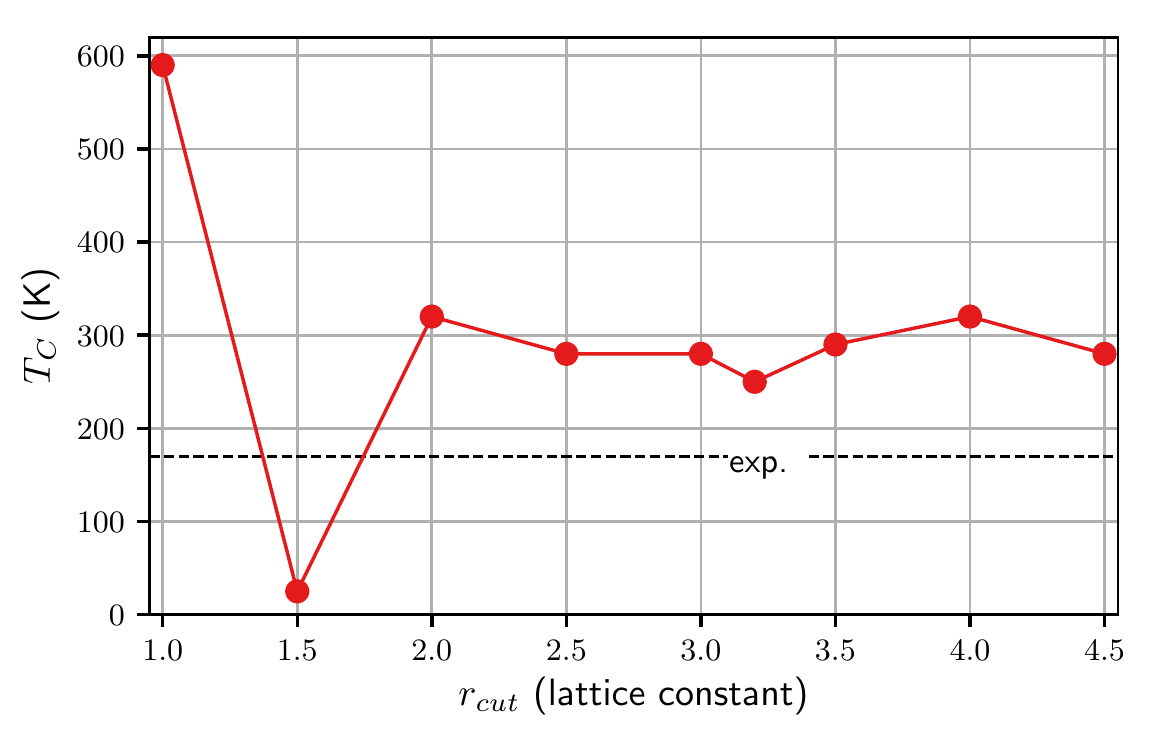}
	\caption{\new{Cutoff distance dependence of the Curie temperature $T_{\text{C}}$ of B20-\ce{MnGe} obtained with a Monte Carlo method.
	The isotropic exchange interactions $J_{ij}$ from \Cref{fig:mnge_varied_u} for $u_\mathrm{Mn}=0.135$ at the theoretical PBEsol lattice parameter were used.}
	%The consideration of interactions above a cut-off distance of $r_{\text{cut}}= 2.0$ does not significantly change the resulting $T_{\text{C}}$.
	}
 \label{fig:mnge_couplings_tc}
\end{figure}
\new{We return to the isotropic pair interactions, $J_{ij}$, and use a Monte Carlo approach to estimate the Curie temperature as a function of the range of the considered interactions.
In this way we can ascertain if the long-ranged nature of these interactions impacts this finite-temperature property.
As a test case, we choose the $u_\mathrm{Mn}=0.135$ case at the theoretical PBEsol lattice parameter.
The results of the Monte Carlo simulations are shown in \Cref{fig:mnge_couplings_tc}.
We find that, although there is a dependence on the number of shells included in the simulations, $T_{\text{C}}$ is more or less converged when interactions up to $2a$ are taken into account, in start contrast to the micromagnetic parameters $A$ and $D$.
This converged value of $T_{\mathrm{C}} \approx \SI{300}{\kelvin}$ lies above the experimental value $T_{\mathrm{C}}^{\mathrm{exp}} \approx \SI{170}{\kelvin}$ \cite{kanazawa_possible_2012}.}

\new{We used the complete set of pair interactions from the previously mentioned dataset, i.e.\ $J_{ij}$ and $\vec{D}_{ij}$, to explore possible magnetic configurations in the lattice spin model.
The atomistic spin dynamics carried out with the Spirit code uncovered the ferromagnetic state as the only stable magnetic structure in 3D simulations, while we could also find metastable skyrmion-like textures when considering thin 2D slabs.
Other than possible numerical difficulties arising from the long-ranged nature of the interactions, the failure of the simulations in finding a stable non-ferromagnetic state in 3D could mean that a model based on bilinear magnetic interactions is insufficient, and that more complex interactions (see e.g.\cite{Brinker2019}) play an important role, as proposed in a recent work\cite{Grytsiuk2019}.
As the first-principles DFT simulations implicitly account for all possible magnetic interactions based on the electronic structure, we next settle whether such non-ferromagnetic structure can be stabilized by performing the appropriate calculations on large supercells of MnGe.}

\section{Large-Scale Electronic Structure Calculations with KKR\lowercase{nano}}
\label{sec:ls_KKRnano_calculations}

\begin{figure*}[tb]
	\centering
	\includegraphics[width=1.0\textwidth]{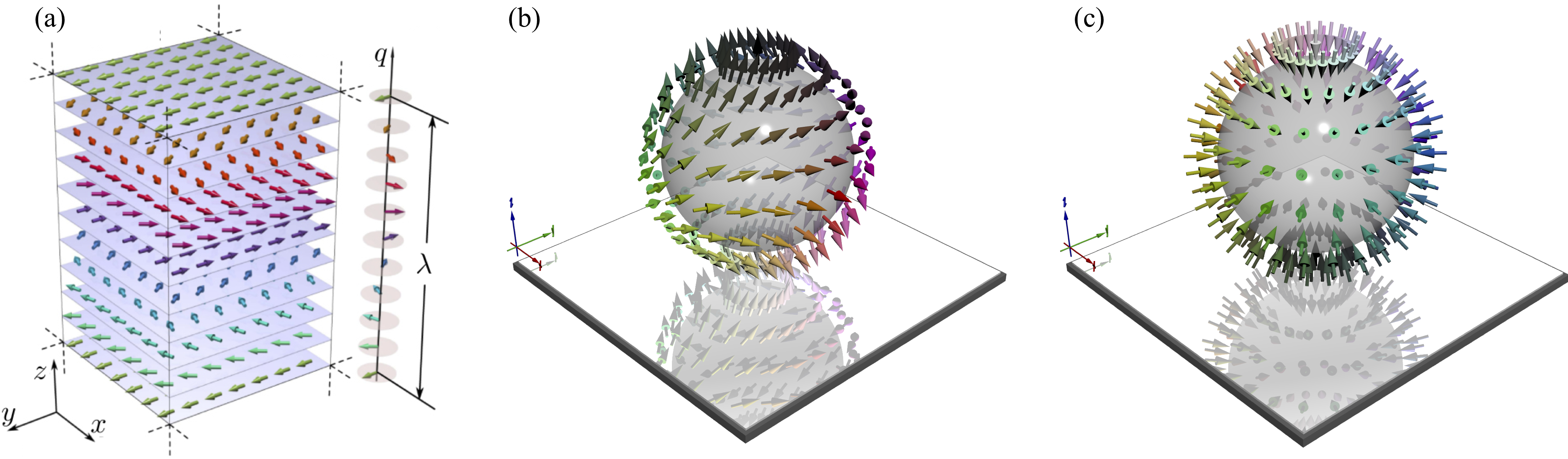}
	\caption{Illustrations of magnetic textures that potentially exist in B20-\ce{MnGe}: (a) Helical spin spiral that propagates in (001) direction.
	(b) 
	Magnetic anti-hedgehog texture that is wrapped around a singularity at the center as it would exist in a 3Q hedgehog-antihedgehog lattice.
   (c) A Bloch point texture where all spins point into the center, thereby creating a monopole.}
\label{fig:mnge_vis_1q_3q_bp}
\end{figure*}

In this section, we present the large-scale DFT results that we obtained with KKR\texttt{nano} for B20-\ce{MnGe}.
\new{We set up a supercell consisting of $6\times6\times6$ conventional unit cells (1728 atoms), which can describe features on a length scale of $6a \approx \SI{3}{\nano\meter}$, the magnetic periodicity found experimentally\cite{kanazawa_large_2011,kanazawa_possible_2012,grigoriev_chiral_2013,shibata_towards_2013,Altynbaev2014,tanigaki_real-space_2015,altynbaev_hidden_2016,kanazawa_critical_2016,kanazawa_topological_2017,altynbaev_magnetic_2018,fujishiro_topological_2019}.}
Besides the previously suggested non-trivial magnetic states, the helical spiral (1Q state) and the hedgehog lattice (3Q state), we explored the possibility of stabilizing the Bloch point (BP) state.
\new{These magnetic structures are illustrated in \Cref{fig:mnge_vis_1q_3q_bp}.}

In our study, the lattice constant is varied and the total energies corresponding to the three states are tracked with respect to the ferromagnetic phase.
As mentioned in the introduction, the main reason that motivates the usage of KKR\texttt{nano} in conjunction with B20-\ce{MnGe} is that Tanigaki et al.\ reported on the existence of the 3Q hedgehog lattice state in this material \cite{tanigaki_real-space_2015}.
Findings by Kanazawa et al.\ suggest that this lattice is set up by a superposition of three orthogonal helical structures also referred to as 3Q state \cite{kanazawa_noncentrosymmetric_2017}. 
Here, the magnetization is determined by the provision
\beq
\label{eq:3q_formula}
\vec{M}(\vec{r}) =
\begin{pmatrix}
	\sin{qy} + \cos{qz} \\
	\sin{qz} + \cos{qx} \\
	\sin{qx} + \cos{qy}
\end{pmatrix},
\eeq
where $q=\frac{2\pi}{\lambda}$ is the wavenumber given in terms of the helical wavelength $\lambda$ and
$x$, $y$ and $z$ are the spatial coordinates within the unit cell.
Note, that $\vec{M}(\vec{r})$ is not normalized. 
\Cref{eq:3q_formula} describes an alternating pattern of hedgehog and anti-hedgehog textures. 

An illustration of an anti-hedgehog is given in \Cref{fig:mnge_vis_1q_3q_bp} b).
Following the micromagnetic description, singularities in the magnetization are expected within the magnetic texture\cite{feldtkeller_continuous_2017}. Our ab initio simulations indicate, however, that all atomic magnetic moments are finite, although our method does not prevent the occurrence of a fully quenched magnetization density within or in between atoms.
In contrast to other systems exhibiting a similar magnetic phase, the rather short helical wavelength of 3 to \SI{6}{\nano\meter} in B20-\ce{MnGe} allows one to perform density functional theory (DFT) calculations with KKR\texttt{nano}.

Other works \cite{makarova_neutron_2012,Deutsch2014a,altynbaev_hidden_2016,Martin2017,yaouanc_magnetic_2017,altynbaev_magnetic_2018} propose that in B20-\ce{MnGe} a helical spin spiral forms along the (001) direction where the magnetization is described by the relation
\beq
\label{eq:1q_spiral}
\vec{M}(\vec{r}) =
\begin{pmatrix}
	 \cos{qz} \\
	-\sin{qz}  \\
	0
\end{pmatrix}
.
\eeq
In the following, we refer to this as the 1Q state (see \Cref{fig:mnge_vis_1q_3q_bp} a)).

Based on our findings in \cref{sec:micro_par}, where we encounter a DM spiralization that does not seem to be larger than \SI{10}{\milli\electronvolt\angstrom}, we also consider a magnetic configuration which can exist without a large DM spiralization but could yield transmission electron microscopy stripe contrasts similar to the 3Q state.
An obvious candidate for this is a Bloch point texture
which can be conveniently defined by means of the four spherical parameters $\phi$, $\theta$, 
$\Phi$ and $\Theta$.
The parameters $\phi$ and $\theta$ designate the position of an \emph{individual atom} in the unit cell
which is described by the common polar and azimuthal angle
\beq
\phi = \arctan{\left(y/x\right)}
\eeq
and
\beq
\theta = \arccos{\left( \frac{z}{\sqrt{x^2+y^2+z^2}} \right)}.
\eeq
Usually, the atomic positions are given in the Cartesian coordinates $x,y$ and $z$.
In the definition above, we define the origin of the coordinate system, i.e. the tuple $(x=0, \, y=0, \, z=0)$,
to be at the center of the unit cell.
In this frame of reference, all atoms that lay in an x-y-plane that intersects with the center
are described by $\theta=\pi/2$.
The orientation of the \emph{individual atomic magnetic moments} for a BP texture is then defined 
by the polar angle
\beq
\Phi = \phi + \phi_{1}
\eeq
and the azimuthal angle
\beq
\Theta = 2 \arctan{\left(\cot{\frac{\theta}{2}} \right)},
\eeq
where the angles designating the atomic position enter as arguments. The phase factor is set to $\phi_{1}=\pi$, hence all magnetic moments point at the origin of the coordinate system.
An illustration of this configuration is given in \Cref{fig:mnge_vis_1q_3q_bp} c).

\begin{figure}[tb]
  \centering
   \includegraphics[width=1.00\columnwidth]{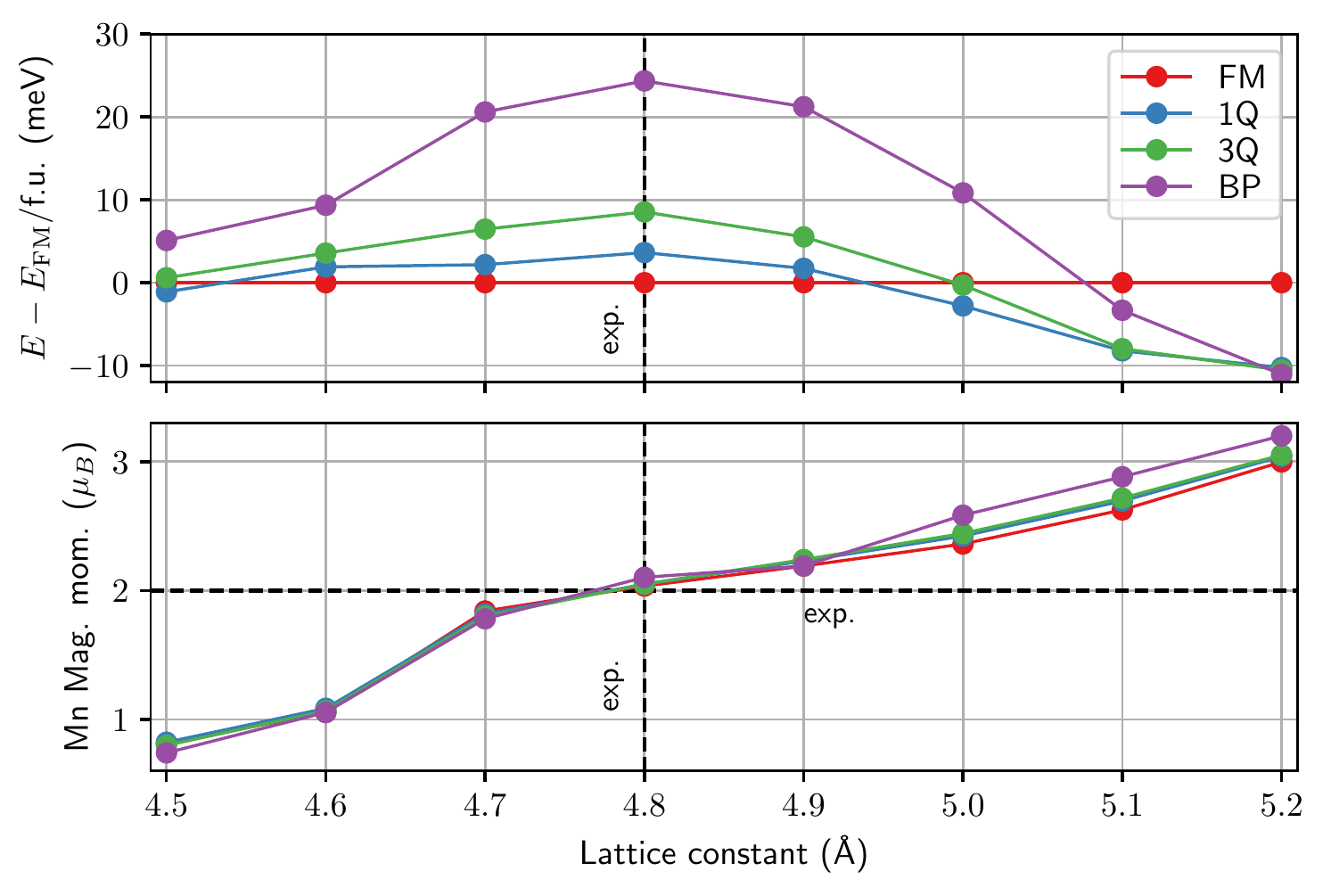}
	\caption{Comparison of Ferromagnetic (FM), helical spiral (1Q), hedgehog lattice (3Q) and Bloch Point (BP)
	state in B20-\ce{MnGe} with KKR\texttt{nano}.
	Top: Difference of total energies with the FM state as reference state for different
	lattice constants. The experimental lattice constant is $a=4.80$ \AA.
	Bottom: Magnetic moment per \ce{Mn} atom increases with lattice constant.
	High-spin/Low-spin transition is clearly visible between $a=4.60$ \AA \, and $a=4.70$ \AA.
	Experimentally, the magnetic moment is measured to be $\approx 2 \mu_B$.}
\label{fig:MnGe_KKRnano_comparison}
\end{figure}

Since the lattice parameter of a material can be modified via strain that 
originates from the manufacturing process of the sample, we investigated the dependence of B20-\ce{MnGe}'s magnetic properties as function of volume.
Such a dependence is depicted in the upper part of \Cref{fig:MnGe_KKRnano_comparison}, where
the total energy is evaluated for FM, 1Q, 3Q and BP states.
The FM state constitutes the ground state,
when the experimental lattice constant is assumed.
However, 1Q and 3Q states are energetically not far from the FM state (within less than \SI{10}{\milli\electronvolt} per Mn atom).
When we further increase the lattice constant the picture changes.
A crucial transition point is found around $a=\SI{5.0}{\angstrom}$, where by imposing the 1Q or 3Q state the energy can be made lower than that of the ferromagnetic state.
In general, for $a>\SI{5.0}{\angstrom}$ 1Q and 3Q states are favored over the ferromagnetic one.
The BP state is energetically not preferred for any lattice constant except for the rather large $a=\SI{5.2}{\angstrom}$.

In the lower part of \Cref{fig:MnGe_KKRnano_comparison} the evolution of the magnetic moment with varying lattice constant is tracked.
The resulting magnetic moment for the experimental lattice constant nicely falls on top of the magnetic moment of approximately $2\,\mu_\mathrm{B}$ per Mn atom which is reported experimentally \cite{makarova_neutron_2012,deutsch_two-step_2014,Altynbaev2014,yaouanc_magnetic_2017}.
The HS/LS transition that is already shown in \Cref{fig:MnGe_hsls} is recognizable between $a=\SI{4.60}{\angstrom}$ and $a=\SI{4.70}{\angstrom}$.
Furthermore, the magnetic moment increases, when the lattice constant is increased.
This is a common behavior which is often observed in metallic systems.
For larger lattice constants the magnetic moments of the different magnetic textures differ more than for the smaller lattice constants.

\section{Conclusions}
\new{We investigated the magnetic properties of B20-\ce{MnGe} through ab initio calculations.
First we considered the basic properties of the ferromagnetic unit cell, recovering the already-known pressure-induced transition from a high-spin to a low-spin state.
Then we analyzed the magnetic interactions derived from the first-principles calculations with both lattice and micromagnetic models.
Lastly, we performed large-scale electronic structure calculations with
KKR\texttt{nano} to quantify the relative energetic stability of different candidate magnetic structures, 1Q (helical spiral), 3Q (hedgehog-antihedgehog lattice) and BP (Bloch point) state, in relation to the ferromagnetic state.
Both the magnetic model simulations and the supercell first-principles calculations found the ferromagnetic state to be the most stable state at the experimental lattice parameter, which is quite close to the theoretically determined one with PBEsol.
This is in clear contradiction with all reported experimental results.}
%We performed full ab initio calculations in order to investigate the magnetic properties of B20-\ce{MnGe}.
%The basic properties of B20-\ce{MnGe} are discussed, in particular the pressure-induced transition from a high-spin to a low-spin state.
%Subsequently, the magnetic exchange interactions that were extracted with the KKR method by means of infinitesimal rotations are presented in \Cref{sec:micro_par}.
%Finally, the energetic behaviour of the magnetic 1Q (helial spiral), 3Q (hedgehog-antihedgehog lattice) and BP (Bloch point) state is analyzed by means of large-scale electronic structure calculations with KKR\texttt{nano}.

\new{The computed magnetic pair interactions showed that the isotropic exchange interactions $J_{ij}$ are very long-ranged and alternate between ferro- and antiferromagnetic as a function of distance, while the Dzyaloshinskii-Moriya interactions $\vec{D}_{ij}$ are rather weak in relation to the former.
This is theoretical evidence for the competing long-range magnetic interactions scenario proposed to explain the short period of the magnetic structure\cite{altynbaev_hidden_2016,altynbaev_magnetic_2018}.
The long-range nature of the magnetic interactions was evidenced by the difficulties found in converging the micromagnetic parameters $A$ (exchange stiffness) and $D$ (spiralization) that can be constructed by performing real-space summations over the corresponding pair interactions.
Nevertheless, the trends in the micromagnetic parameters are in good correspondence with previous theoretical works\cite{gayles_dzyaloshinskii-moriya_2015,kikuchi_dzyaloshinskii-moriya_2016,mankovsky_composition-dependent_2018,Mankovsky2019}.
We envision two scenarios that could explain the experimental findings considering only pair interactions, both invoking competing exchange interactions: (i) $A \approx 0$, so that a weak $D$ can stabilize helical modulations; (ii) $A < 0$, signaling the instability of the ferromagnetic state to helical modulations, with $D$ playing a secondary role of selecting the chirality of the modulation.}

\new{The first-principles calculations performed for the ferromagnetic, 1Q, 3Q and Bloch point structures found that it is possible to stabilize either the 1Q or the 3Q structures with respect to the ferromagnetic state if the lattice parameter is increased.
This can be interpreted by the computed magnetic pair interactions, which showed a change in sign of $A$ due to a weakening of the ferromagnetic nearest-neighbor $J_{ij}$, and a strengthening of $D$ due to a rotation of the $\vec{D}_{ij}$ towards the bond directions, which more than compensates the reduction in their magnitudes.
Overall, the energy differences between the 1Q and 3Q structures and the ferromagnetic state was at most \SI{10}{\milli\electronvolt} per Mn atom, which is a model-independent verification that the magnetic interactions in the system are indeed competing, so that quite different magnetic structures have very similar energies.
Regarding the Bloch point structure, the imposition of periodic boundary conditions in KKR\texttt{nano} means that the magnetic spins at the boundaries of each supercell are aligned in an unfavorable antiferromagnetic way and thus there is a large energy penalty.
Notably, this energy cost decreases by increasing the lattice parameter, which can again be related to the reduction of the nearest-neighbor $J_{ij}$.}

\new{As all our results are based on first-principles calculations, a possible explanation for the disagreement with experiment could be in the computed electronic structure.
First, the small energy differences found for the supercell calculations might make the results sensitive to intrinsic deficiencies of the exchange-correlation functional.
Second, and perhaps more likely, could be an excessive delocalization of the Mn $d$-orbitals.
This is a well-known common failure of the standard functionals, which could upset the energetic balance between the different magnetic structures by strengthening the nearest-neighbor $J_{ij}$.
The work reported in Ref.~\cite{Choi2019} could be interpreted in this way, and the authors do find several noncollinear magnetic structures by tuning the coupling between itinerant and more localized electrons.}

%Our first-principles based simulations indicate that neither the experimentally reported 1Q nor the 3Q texture is predicted to be the ground state when assuming the equilibrium lattice constant.
%This contradicts the reports from experiment.
%Nevertheless, our results clearly show that B20-\ce{MnGe} features strongly competing magnetic interactions which means that different magnetic states can have very similar energies.
%At this point, it must be considered that an exchange-correlation functional like PBEsol might not provide total energies that are precise enough to correctly capture such subtle differences.
%However, by increasing the lattice constant both 1Q and 3Q texture can be made energetically preferable.
%A possible explanation for this is the first-nearest neighbor isotropic magnetic exchange coupling which, as shown in \Cref{fig:mnge_varied_alat}, is lowered for larger lattice constants.

Experimentally, it is not possible to have the ideal lattice structures considered in our study. Thus various effects can affect the experimental observations and the related interpretations. For instance, impurities in the sample can potentially exert chemical pressure, which leads to spatial expansion of the
lattice structure (see the example of \ce{Co}-doped B20-\ce{FeGe} \cite{stolt_chemical_2018}). \new{Overall our study motivates further theoretical and experimental investigations of three-dimensional magnetism in B20 materials in general and in  B20-\ce{MnGe} in particular. }
%\mdsd{I think that we should cut this paragraph, as there are plenty of experiments by different groups arriving at the same results, at least concerning the helical modulations, and there have been careful studies of the structure and chemical composition of the samples too.}{\color{red}Samir: I am not against rising the question that the experimental interpretation of the 3D structures requires further investigations.}

%Regarding the BP texture, it should be stressed that in KKR\texttt{nano} periodic boundary conditions are assumed and therefore the magnetic spins at the edges of each periodic image are aligned in a non-favorable antiferromagnetic way.
%Thus there is a large energy-penalty to pay.
%Interestingly, by increasing the lattice parameter, this energy cost decreases which can again be related to the reduction of the first-nearest neighbor isotropic exchange interaction upon increase of the lattice parameter.

\section*{Acknowledgements} 
We thank Nikolai S. Kiselev for fruitful discussions.
We  acknowledge funding from Deutsche  For\-schungsgemeinschaft (DFG) through SPP 2137 ``Skyrmionics", the Collaborative Research Centers SFB 1238 and SFB/TRR 173. S.B.\ acknowledges the DARPA TEE program through grant MIPR\# HR0011831554 from DOI. S.L. and M.d.S.D. acknowledge support from the European Research Council (ERC) under the European Union's Horizon 2020 research and innovation program (ERC-consolidator grant 681405 -- DYNASORE). We gratefully acknowledge  the Gauss Centre for Supercomputing e.V. (www.gauss-centre.eu) for funding this project by providing computing time through
the project GCS-KKRN on the GCS Supercomputer Hazel Hen at Höchstleistungsrechenzentrum Stuttgart (HLRS). We further acknowledge Forschungszentrum Jülich/Jülich Supercomputing Centre for granting access to the QPACE 3 computer, which was built as part of the DFG (SFB/TRR 55) project.

\bibliography{mylit}% Produces the bibliography via BibTeX.

\appendix
\section{Connection between Atomistic and Micromagnetic Model in B20 Materials}
\label{sec:con_atom_micro}
The atomistic model parameters, which appear in the well-known Heisenberg model, can be connected to 
the micromagnetic model, that has the form of a continuum theory.
The latter is widely used in the skyrmion community and we adopt it to complement our
toolbox for the investigation of the magnetic properties of B20-\ce{MnGe}.
We exemplify the connection between atomistic and continuum model by considering a helical spin spiral that points along the z-axis
(c.f. 1Q state in \cref{eq:1q_spiral}) 
and is described by
the wave vector $\vec{q}= \begin{pmatrix} 0, 0, q \end{pmatrix}$, i.e. the magnetic moments rotate
within the x-y-plane and the wave vector points along the z-axis.
The magnetization of each atom $i$ is then given by
\beq
\label{eq:spinspiral_z}
\vec{m}_i = \cos{(qz_i)} \hat{e}_x - \sin{(qz_i)} \hat{e}_y,
\eeq
where $z_i$ denotes the z-coordinate of the respective atomic site. It can be shown that such a magnetic structure interpolates smoothly between the discrete lattice and the continuum limit.
We define the Heisenberg energy with isotropic exchange interaction
and DM interaction as
\beq
E_{\text{atom}} = -\sum_{ij} J_{ij} \vec{m}_i \vec{m}_j +
\sum_{ij} \vec{D}_{ij} \cdot \left(\vec{m}_i \times \vec{m}_j \right).
\eeq
Insertion of \cref{eq:spinspiral_z} and usage of addition theorems leads to
\begin{align}
	\label{eq:h_atom1q}
	E_{\text{atom,1Q}} %&= -\sum_{ij} J_{ij} \cos{(q(z_i-z_j))} \nonumber \\ 
	%&+ \sum_{ij} \vec{D}_{ij} \cdot \hat{e}_{z} (\sin{(qz_i)} \cos{(qz_j)} \nonumber \\
	%&- \cos{(qz_i)} \sin{(qz_j)}) \nonumber \\
	&= -\sum_{ij} J_{ij} \cos{(q(z_i-z_j))} \nonumber \\
	&+ \sum_{ij} D_{ij}^{z} \sin{(q(z_i-z_j))} \nonumber \\
	&= N \left(-J(q) + D^{z}(q)\right),
\end{align}
where we used the translational invariance of $J_{ij}$ and $N$ is the number of atoms.
For the helical spiral defined in \cref{eq:spinspiral_z}, only the z-component $D_{ij}^{z}$ of $\vec{D}_{ij}$
needs to be considered.

The micromagnetic energy reads
\begin{align}
E_{\text{micro}} =& \int dV\, A \left((\nabla m_x)^2 +  (\nabla m_y)^2 + (\nabla m_z)^2 \right) \nonumber \\
&+ D \vec{m} \cdot \left(\nabla \times \vec{m} \right),
\end{align}
where $A$ is the so-called \textit{spin stiffness} and $D$ the \textit{DM spiralization} \cite{nagaosa_topological_2013}.
Insertion of
the magnetization of the helical spiral given by \cref{eq:spinspiral_z} yields
\begin{align}
	E_{\text{micro},1Q} &= \int\mathrm{d}V\,\bigg[A q^2  %\left(\sin^2{(qz)} + \cos^2{(qz)}  \right) \nonumber \\
	+ D \left(\cos{(qz)} \hat{e}_x - \sin{(qz)} \hat{e}_y \right) \cdot \nonumber \\
	&\hspace{4em}\cdot \left(\frac{\partial}{\partial z}  \sin{(qz)} \hat{e}_x
	+ \frac{\partial}{\partial z} \cos{(qz)} \hat{e}_y \right)\bigg] \nonumber \\
	%=&  \int dV\, A q^2 + Dq  \left(\cos^2{(qz)}  + \sin^2{(qz)} \right) \nonumber \\
	%=&  \int dV\, A q^2 + Dq  \nonumber \\
	&= A q^2 + Dq.
\end{align}
The wave number $q$ will take the value which minimizes $E_{\text{micro},1Q}$ and we can thus
impose the condition
\begin{align}
	\label{eq:helical_pitch}
	\frac{\partial E_{\text{micro},1Q}}{\partial q} \stackrel{!}{=} 0 %\nonumber \\
	%\Leftrightarrow \, &  \frac{\partial}{\partial q}
	%\left(A q^2 + Dq \right)=0 \nonumber \\
	%\Leftrightarrow \, & 2Aq + D  =0 \nonumber \\
	\Leftrightarrow\,  q  = -\frac{D}{2A},
\end{align}
which gives us a provision on how the wave number $q$ depends on the magnitude of DM spiralization
and spin stiffness.

The atomistic and the micromagnetic model are connected in the limit $q \rightarrow 0$, i.e. for a helical spiral
that extends over multiple unit cells. \Cref{eq:h_atom1q} can then be simplified to
\begin{align}
	\label{eq:tot_energy_micro}
	E_{\text{atom,1Q}}=& -\sum_{ij} J_{ij} \cos{(q(z_i-z_j))}
	+ \sum_{ij} D_{ij}^{z} \sin{(q(z_i-z_j))} \nonumber \\
	=&  -\sum_{ij} J_{ij} \left(1-\frac{1}{2}\left(q(z_i-z_j) \right)^2\right) \nonumber \\
	&+ \sum_{ij} D_{ij}^{z} q(z_i-z_j) + \mathcal{O}(q^3) \nonumber \\
	=&  \underbrace{-\sum_{ij} J_{ij}}_{E_0} + \underbrace{\frac{1}{2}\sum_{ij} J_{ij} \left(z_i-z_j \right)^2}_{A}q^2 \nonumber \\
	&+ \underbrace{\sum_{ij} D_{ij}^{z} \left(z_i-z_j \right)}_{D} q + \mathcal{O}(q^3).
\end{align}
Thus in this limit,
it is possible to derive the micromagnetic parameters $A$ and $D$ from the atomistic parameters $J_{ij}$
and $\vec{D}_{ij}$ which can be obtained from a KKR calculation by following the procedure described in 
\cite{liechtenstein_local_1987,udvardi_first-principles_2003,ebert_anisotropic_2009}. The term $E_0$ determines the ferromagnetic reference energy.
The exchange stiffness $A$ describes the increase in energy if a spin spiral is assumed instead of the
ferromagnet. The micromagnetic DMI $D$ can lower the energy if the product of $D$ and $q$ is negative
and can thus make the spin spiral configuration the energetically preferred state.

In general, $A$ and $D$ are $3 \times 3$ tensors that we denote with $\mathcal{A}$ and $\mathcal{D}$.
For B20 compounds
this simplifies to diagonal matrices
due to symmetry arguments and we obtain
\begin{align}
	\label{eq:stiff_tensor}
	\mathcal{A}
	&= \frac{1}{4} \sum_{\text{s}} J_{s}
\begin{pmatrix}
	 4\vec{R}_{s} \cdot  \vec{R}_{s} & 0 & 0 \\
	0 & 4\vec{R}_{s} \cdot  \vec{R}_{s}  & 0 \\
	0 & 0 &  \vec{R}_{s} \cdot  \vec{R}_{s}
\end{pmatrix} \nonumber
\\
	&=
	 \sum_{s} J_{s} \left|\vec{R}_{s}\right|^{2} \mat{I_3}
= A \mat{I_3}
\end{align}
and
\begin{align}
	\label{eq:spir_tensor}
	\mathcal{D}
	&=  \frac{1}{2} \sum_{s}
\begin{pmatrix}
	4\vec{R}_{s} \cdot  \vec{D}_{s} & 0 & 0 \\
	0 & 4\vec{R}_{s} \cdot  \vec{D}_{s}  & 0 \\
	0 & 0 &  4\vec{R}_{s} \cdot  \vec{D}_{s}
\end{pmatrix} \nonumber
\\
	&= 2
	\sum_{s} 
	\left(\vec{R}_{s} \cdot \vec{D}_{s}\right) \mat{I_3}
	= D \mat{I_3}.
\end{align}
Here, the summation is performed over all shells $s$, so that symmetrically equivalent parameters are omitted.
It should be noted that from \cref{eq:spir_tensor} it follows that $D$ vanishes for
$\vec{D}_n \perp \vec{R}^{n}$ and is largest for $\vec{D}_n \parallel \vec{R}^{n}$.

\end{document}